\documentclass[12pt,a4paper]{article}
\usepackage{hyperref}
\usepackage{siunitx}
\usepackage{mathptmx}
\usepackage{graphicx} 
\usepackage{soul} %for consistant underlining \ul{}
\usepackage{float}

\usepackage{amsmath}
\usepackage{amsfonts}
\usepackage{amssymb}
\usepackage[a4paper, left=25mm, right=25mm, top=25mm, bottom=20mm]{geometry}
\begin{document}

\begin{titlepage}

{\flushright{
        \begin{minipage}{2.5cm}
         DESY 20-027
        \end{minipage}        }

}
%Titel

\begin{center}
{\LARGE\bf
Plasma Lenses: Possible alternative OMD at the ILC\footnote{Talk presented at the International Workshop on Future Linear Colliders (LCWS2019), Sendai, Japan, 28 October-1 November 2019.}}
\vskip 1.0cm
{\large Manuel Formela$^a$,
Niclas Hamann$^a$, Klaus Fl\"ottmann$^b$, 
        Gudrid Moortgat-Pick$^{a,b}$, Sabine Riemann$^c$, Andriy Ushakov$^{a}$}
        \vspace*{8mm}\\
{\sl\small 
${}^a$ University of Hamburg, Luruper Chaussee 149, D-22761 Hamburg, Germany \\
${}^b$ Deutsches Elektronen-Synchrotron (DESY), Notkestraße 85, D-22607 Hamburg, Germany\\
${}^c$ Deutsches Elektronen-Synchrotron (DESY), Platanenallee 6, D-15738 Zeuthen, Germany}

\end{center}
%\date{}

\begin{abstract}
In the baseline design of the International Linear Collider (ILC) an undulator-based source is foreseen for the positron source in order to match the physics requirements. The recently chosen first energy stage with $\sqrt{s}=250$~GeV requires high luminosity and imposes an effort for all positron source designs at high-energy colliders. In this paper we perform a simulation study and adopt the new technology of plasma lenses to capture the positrons generated by the undulator photons and to create the required high luminosity positron beam.
\end{abstract}

\end{titlepage}

\section{Motivation}
The International Linear Collider (ILC) as well as the multi-Tev high-energy collider design CLIC have to provide 
polarized beams at high intensity as well as at high energy. Challenging is the production of the high-intense positron beams.
The ILC uses an undulator-based positron source in the baseline design~\cite{Adolphsen:2013kya,ilc-rdr,BCD} that even produces a polarized positron beam. In this way, i.e. offering both high intense and highly polarized electron and positron beams, the physics potential of the ILC is optimized and well prepared for high precision physics as well as for any new 
discoveries~\cite{Moortgat-Picka:2015yla,AguilarSaavedra:2001rg}. Currently an initial energy of $\sqrt{s}=250$~GeV is discussed~\cite{Fujii:2018mli}, where the undulator scheme
can be applied as well~\cite{Ushakov:2018wlt,Sabine}. 
It has been shown that the physics precision requirements can not be fulfilled if only polarized electrons were available since
in that case the systematic uncertainties get too large, see \cite{Robert-Thesis,Karl:2017xra,Aihara:2019gcq}. 

Since the luminosity requirements are challenging  for any kind of positron sources at a high-energy colliders we study in the following  a new idea, 
namely using a plasma lens (PL) as optic matching device instead of the commonly used quarter-wave transformer (QWT). 
The use of a plasma lens as an optical matching device (OMD) at the positron source  is a novel application 
with a high potential to improve the yield and the quality of the positron bunch.
We work out first design parameters of such a PL and compare the results with those from a QWT~\cite{Adolphsen:2013kya,ilc-rdr,2,3}. 

%We are interested in the use of a plasma lens as an optical matching device (OMD) at ILC, i.e. for focussing the divergent positron bunch emerging from the target and matching its properties to the damping ring acceptance. Currently the quarter wave transformer (QWT) and flux concentrator (FC) are designed and simulated by M. Fukuda to function as an OMD at ILC \cite{ilc-rdr}. 
%The problem with these more conventional approaches to the optical matching of charged particle bunches is 
%
%The use of a plasma lens as an OMD is a novel application 
%with potential to improve yield and quality of the positron bunch, because the PL offers a azimuthal magnetic field leading to a radially symmetric focussing.

\section{Current Design: Quarter-wave transformer}

\subsection{Status QWT at ILC with $\sqrt{s}=250$~GeV}

\begin{figure}[H]
	\centering

	\includegraphics[scale=.5]{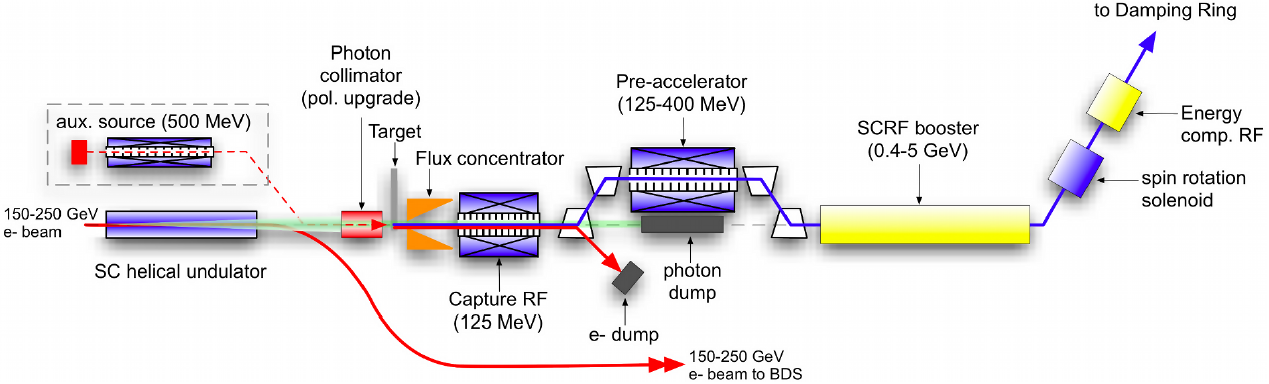} 
    \caption{Schematic layout of positron source \cite{Adolphsen:2013kya}}
    \label{fig:positron source layout}
\end{figure}

In this section we simulate the current scheme, using a quarter-wave transformer as OMD, with our tools and compare the results with results from \cite{2,3}. 

Figure~\ref{fig:positron source layout} shows the schematic layout of the ILC positron source including the production, capture and transfer of the positron beam: electron bunches accelerated to energies between $125-\SI{250}{\GeV}$ are injected into the helical undulator. The helical undulator is a periodic arrangement of electromagnets designed to force the electrons onto a helical trajectory, irradiating circularly polarized photons into a cone in forward direction due to the transverse accelerating process. While the electron bunch is redirected to the beam delivery system (BDS), the radiated photons are  led through a photon-collimator ---increasing the mean polarization--- onto a rotating
%\SI{7}{\cm}$ thick,
Ti-6\%Al-4\%V target. 
Inside the target electroweak interactions between atoms and photons with energies above a lower limit result in pair production of longitudinally polarized electrons and positrons. Only the positrons are kept; the parameters of these still divergent positrons are required to be matched to the acceptance requirements of the downstream damping ring. This matching can be achieved with the optical matching device (OMD), which in this case is currently foreseen to be  either a quarter-wave transformer (QWT) or a flux concentrator (FC). The OMD is followed by the capture RF cavity, which accelerates the positron bunch to $\SI{125}{\MeV}$. Further downstream elements before the damping ring are the electron and photon dump, SCRF booster, spin rotation solenoid, energy compression structure, cf. Figure~\ref{fig:positron source layout}~\cite{1,Adolphsen:2013kya}.

\subsection{Comparison of the results for the QWT}

The QWT is a solenoid encased in a tapered iron shell used to capture the divergent positron shower leaving the target and  presents the heart  of the OMD. The shell is needed to minimize eddy currents induced in the rotating target by the magnetic field of the OMD.\\

Before simulating the various plasma lenses designs as OMD  and examine their effectiveness, efficiency, limits, etc. as an OMD for the ILC, we 
use the program  ASTRA \cite{5} for simulating the QWT device. We benchmarked our simulations on a QWT design, discussed and simulated in GEANT4 \cite{2,3}, for the following three reasons:
\begin{itemize}
	\item verification of the current QWT simulations~\cite{2,3};
	\item understanding of existing analyses of various positron beam properties;
	\item obtaining reference results for plasma lens originating from the same simulation program (ASTRA).
\end{itemize}

\begin{figure}[H]
	\centering
	\includegraphics[scale=.55]{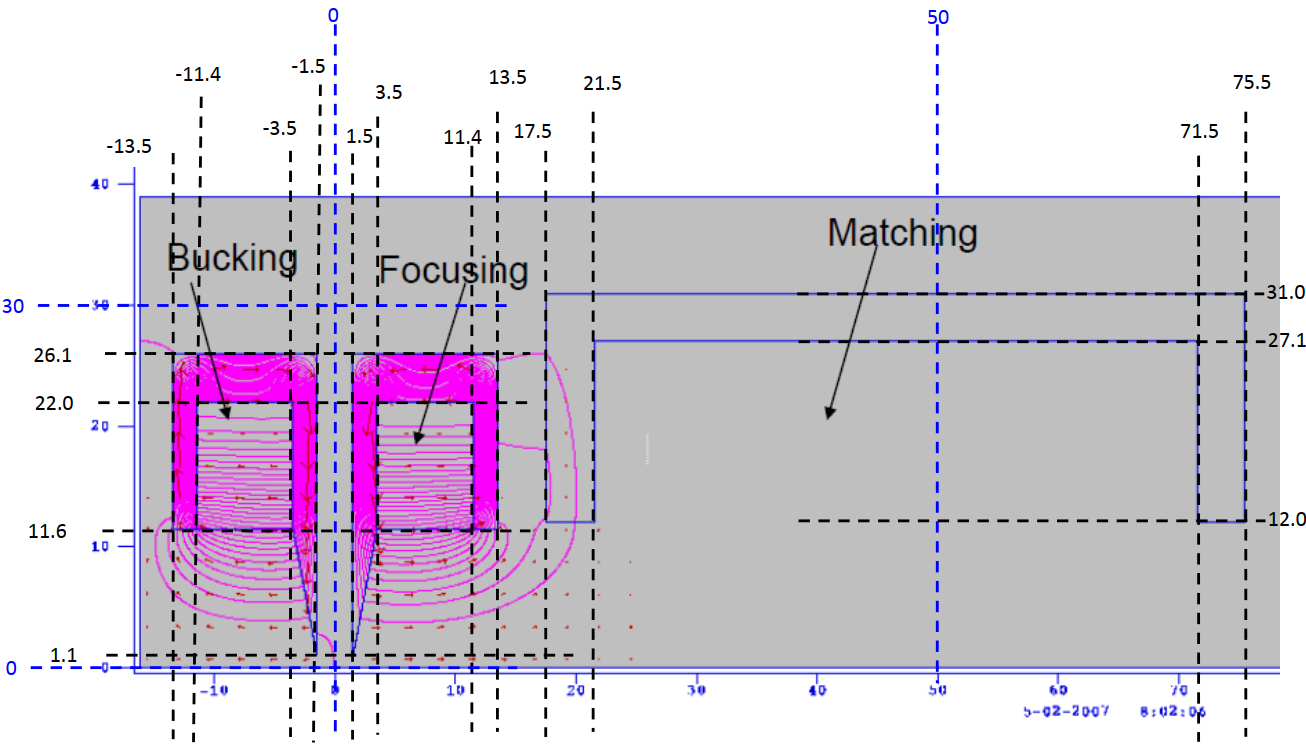} 
    \caption{Geometry of the simulated QWT (focusing solenoid) nestling between the bucking and matching solenoid (lengths in cm) \cite{2}.}
    \label{fig:QWT geometry}
\end{figure}

In the following, the QWT geometry \cite{2} and its magnetic field used for the QWT-simulation results below are described.
The QWT is located $\SI{7.6}{\mm}$ from the rear side of the target and $\SI{40}{\mm}$ from the front side of the RF cavity solenoid~\cite{3}. Its iron casing's front side opening is $\SI{1.1}{\cm}$ in radius and is linearly tapered to $\SI{11.6}{\cm}$ over a distance of $\SI{2}{\cm}$, followed by two sections of constant radii of $\SI{22}{\cm}$ over $\SI{7.9}{\cm}$ and $\SI{11.6}{\cm}$ over $\SI{2.1}{\cm}$, respectively, where the solenoid itself sits within the former (see figure \ref{fig:QWT geometry})\cite{2}.\\

\begin{figure}[H]
	\centering
     \includegraphics[scale=.7]{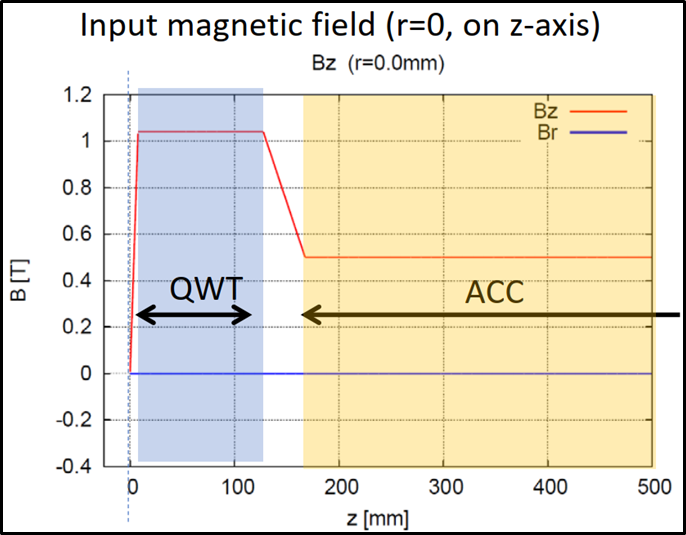} 
    \caption{The approximated on-axis longitudinal magnetic field of the QWT including the solenoid from the cavity section~\cite{3}.}
    \label{fig:approximated QWT B-field}
\end{figure}

The on-axis longitudinal magnetic field $B_z(r=0)$ of the QWT ---,while also taking the solenoid along the RF cavities into account,--- can be approximated by sections of constant, linearly increasing and linearly decreasing magnetic fields (see figure \ref{fig:approximated QWT B-field}). The field 
starts with $\SI{0}{\tesla}$ at the target's rear side, then increases linearly to $\SI{1.04}{\tesla}$ at the front side of the QWT. A constant field of 
$\SI{1.04}{\tesla}$ is assumed to approximate the field across the whole QWT length, followed by a section of linear decrease to $\SI{0.5}{\tesla}$ at the front side of the RF cavity. The final bit of the magnetic field, located across the RF cavity, is constant and achieves $\SI{0.5}{\tesla}$~\cite{3}.\\
Declaring only the on-axis longitudinal magnetic field values is sufficient to run the simulations, because the radial dependency as well as the radial magnetic field component can be derived from a polynomial expansion~\cite{6}. 

The QWT simulation, presented now, does not consider space charge effects , i.e. the Coulomb interaction within the positron bunch. It has been assumed, however,  that the positrons are only affected by the QWT'S magnetic field and by its geometry.
The used positron distribution, has been derived from GEANT4 simulations based on undulator radiation simulation in CAIN\cite{Fukuda-Yokoya,3}.
The obtained  results are: from the initial $42917$ positrons exiting the target  $33732$ pass the QWT in our simulation  and $35865$  in those of \cite{Fukuda-Yokoya}, respectively. This is a difference of less than $6\%$. Almost all lost positrons enter the front side of the QWT.

Comparing now the positron distributions in terms of their transversal momentum $p_t$, divergence $\sin \theta = p_t/p_z$ and energy $E$ reveals good agreement between both simulation approaches. The observables were always taken for every single positron at the entrance of the RF cavity solenoid, exactly $\SI{40}{\mm}$ away from the QWT's rear side,  cf.\ the case that the positron beam was dumped onto a target exactly  at this position.
%While comparing the plot results, we neglect the red curve simulated by A. Ushakov and focus only on the blue curve by M. Fukuda.

\begin{figure}[H]
	\centering
    \includegraphics[scale=.7]{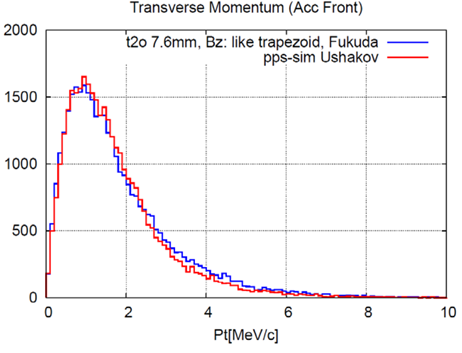}
    \includegraphics[scale=.7]{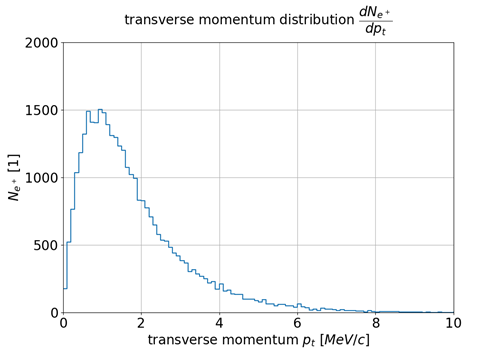} 
    \caption{Comparison of the transversal momentum distribution of the positron bunch at the entrance of the cavity solenoid (right panel)  with
    the corresponding results from \cite{3} (left panel).}
    \label{fig:Ne vs pt}
\end{figure}

In figure~\ref{fig:Ne vs pt} the simulated
momentum distribution $dN_{e^+}/dp_t$ with a bin size of $\SI{0.1}{\mega \electronvolt \per c}$ is plotted (right panel) and compared with the corresponding result from \cite{3} (left panel). Both graphs have very similar trends, deviating only on the peak size, which can be explained by the previously stated $6\%$ difference in the total number of positrons.

\begin{figure}[H]
	\centering
	\includegraphics[scale=.7]{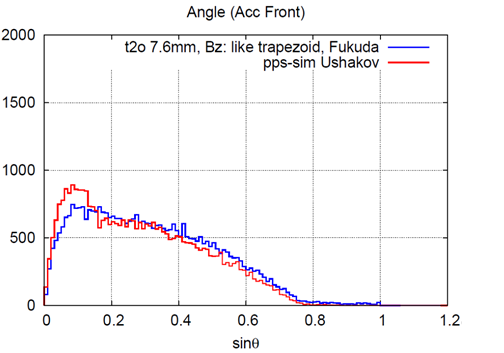} 
    \includegraphics[scale=.7]{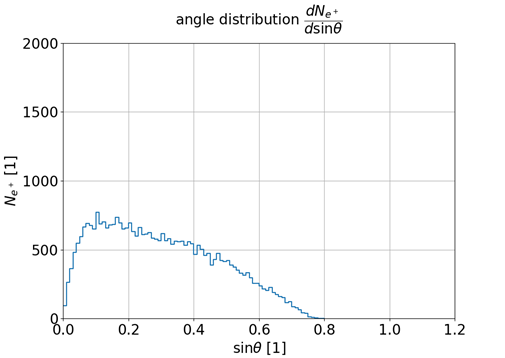}
    \caption{Comparison of the divergence distribution of the positron bunch at the entrance of the cavity solenoid (right panel)  with
    the corresponding results from \cite{3} (left panel).}
    \label{fig:Ne vs sin theta}
\end{figure}

In figure \ref{fig:Ne vs sin theta} the derived divergence distribution $dN_{e^+}/d\sin \theta$ with a bin size of $0.01$ is plotted (right panel)
and compared with the corresponding result from \cite{3} (left panel).
Again the two graphs are very similar.
%, but this time the decrease is to be found at the back end of the curve, suggesting less positrons with high divergence.

\begin{figure}[H]
	\centering
    \includegraphics[scale=.7]{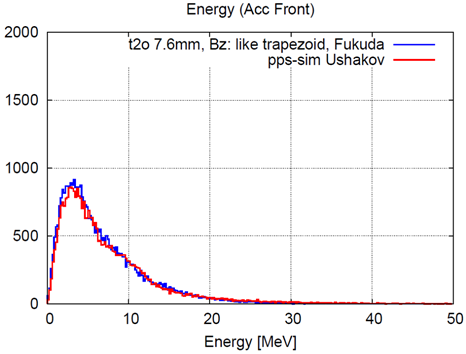}
    \includegraphics[scale=.7]{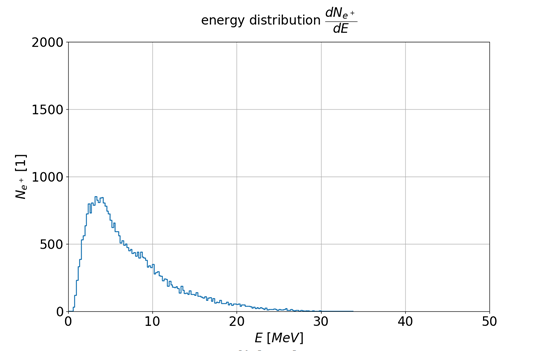} 
    \caption{Comparison of the energy distribution of the positron bunch at the entrance of the cavity solenoid (right panel)
     with the corresponding results from \cite{3} (left panel).}
    \label{fig:Ne vs E}
\end{figure}

Finally,  the energy distribution $dN_{e^+}/dE$ with a bin size of $\SI{0.2}{\mega \electronvolt}$ is plotted in figure \ref{fig:Ne vs sin theta} (right panel)
and compared with the corresponding result from \cite{3} (left panel).
Concerning this energy distribution both graphs are nearly indistinguishable.
% likely due to the fact that the bin size is smaller compared to the whole plot width, making the differences less apparent.\\

Concluding our QWT simulation section, one can state that the code ASTRA has been successfully be applied to simulate the OMD, cf. \cite{3}. 
 In the following, we therefore also use ASTRA as reference simulation for the plasma lenses. 
 %More work on the QWT simulation could be useful for example when analysing the phase space or even adding the RF cavities.

\section{Novel idea: active plasma lenses}
\subsection{Fundamental principles}
\begin{figure}[H]
	\centering
    \includegraphics[scale=.95]{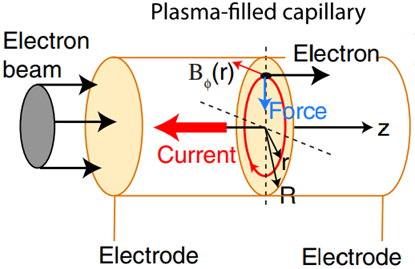}
    \includegraphics[scale=.75]{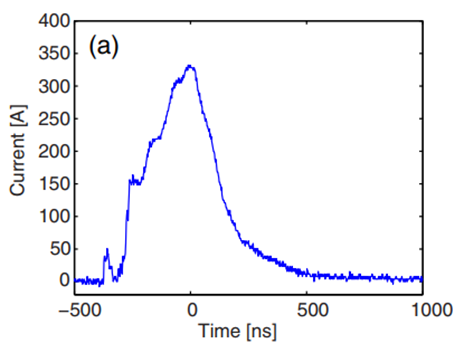}
    \caption{The principle of an active plasma lens (left panel) and the characteristic discharge current pulse in plasma lenses (right panel) \cite{vanTilborg}.}
    \label{fig:PL scheme and current pulse}
\end{figure}

The active plasma lens (PL) set-up consists of a capillary, filled with gas (e.g. $H_2$), having an electrode  each attached to both opposing openings, see figure~ \ref{fig:PL scheme and current pulse} (left panel) \cite{vanTilborg}. A pulser system supplies both cathode and anode with a short multi-$\SI{}{\kilo \volt}$ voltage pulse. The strong electric field along the capillary leads to the ionization of the gas, i.e. freeing electrons from bondings with atoms/molecules and therefore forming a gas mixture of free electrons and positively charged ions, called a plasma. 
Furthermore the electric field also accelerates the free electrons in the direction of the cathode leading to a strong sub-$\SI{}{\micro\s}$ axial discharged current pulse in the order of up to some hundreds of Ampere, see figure~\ref{fig:PL scheme and current pulse} (right panel). The moving charge induces an azimuthal magnetic field ---similar to a wire--- which then in turn allows for a radially symmetric focussing of charged particle beams passing the 
capillary and accordingly also the plasma with negligible  interaction. 

One should note that any kind of windows, that would be vulnerable to drastic stress from the positron beam, are obsolete in the case at hand due to the use of differential pumping of the capillary gas.\\

\subsection{Design results}

In order to get a parallel beam from such a rather divergent positron beam, mentioned above, one has to use an decreasing magnetic field along the z-axis,
applying, for instance, such tapered plasma lenses. These lenses vary their radius along the z-axis, which results in altering the magnetic field. 
These lenses belong to the class of adiabatic matching device (AMD) for focusing electron/positron beams.
 It is the specific manner how the magnetic field decreases that made tapering of plasma lenses 
 appear to be the right approach to apply PLs as an OMD\cite{7}.
 %The reason is the special shape in which the magnetic field decreases. 
%This made the tapering of a plasma lenses appear to be the right approach to utilize a PL as an OMD\cite{7}. 
The basic elements of future designs will consist of a tapered and a constant plasma lens. This combination of tapered plasma lens followed by a constant plasma lens will be referred as device.\\

With regard to the OMD for the  positron source of the ILC we finally came up ---after varying many parameters--- with a first design proposal. 
It consists of a down tapered plasma lens and a constant lens, which was previously demonstrated experimentally to work in principle as one integrated device\cite{7}. Both lenses are $3\,\mathrm{cm}$ long and the constant one immediately starts after the tapered plasma lens. The tapered plasma lens starts at $z=7.6\,\mathrm{mm}$ with a radius of $10\,\mathrm{mm}$ which expands linearly to $25\,\mathrm{mm}$ within $30\,\mathrm{mm}$. 
The constant one has the radius $25\,\mathrm{mm}$. All in all the whole device is only $60\,\mathrm{mm}$ long and the current strength is set to $2500\,\mathrm{A}$, cf. Table~\ref{tab:Tab2}, more details see \cite{niclas-bachelor}. \\

\begin{table}[H]
	\begin{center}
		\begin{tabular}{ |c|c| } 
			\hline
			Total number of particles on stack & 1000 \\ 
			Positrons & 1000 \\ 
			particles at the cathode & 0 \\ 
			active particles & 766 \\ 
			passive particles (lost out of bunch) & 0 \\ 
			probe particles & 0 \\ 
			backward traveling particles & 28 \\ 
			particles lost with z$<$Zmin & 0 \\ 
			particles lost due to cathode field & 0 \\ 
			particles lost on aperture & 234 \\ 
			\hline
		\end{tabular}
	\end{center}
	\caption{Plasma lens as design for the OMD for the ILC positron source (more details, see \cite{niclas-bachelor}).}
	\label{tab:Tab2}
\end{table}

In our simulation with the ASTRA code, the space charge is neglected in first approximation and only the interaction between particle and magnetic field has been taken into account, similar as in the previous case when benchmarking the code with the simulation for the QWT. 

The following beam statistics are calculated up to $15\,\mathrm{mm}$ after the device ended, i.\ e.\ at the final position $z=9.26\,\mathrm{cm}$. As can be seen in table \ref{tab:Tab2} $766$ positrons from the initial $1000$ simulated positrons are still active at this point, meaning they were not lost on the plasma lens geometry. It is important to note, however, that some of the active particles also travel backwards. This decreases the effective active particle count slightly to $738$ positrons when measuring  the statistics.\\

\begin{table}[H]
	\begin{center}
		\begin{tabular}{ |c|c| } 
			\hline
			Particles taken into account N & 766\\ 
			total charge  Q& 1.22E-07 nC \\  
			horizontal beam position x & 0.15 mm \\ 
			vertical beam position y & 0.31 mm \\ 
			longitudinal beam position z & 9.26E-02 m \\ 
			horizontal beam size sig x & 15.71 mm \\ 
			vertical beam size sig y & 16.92 mm \\ 
			longitudinal beam size sig z & 25.68 mm \\ 
			average kinetic energy E & 6.78 MeV \\ 
			energy spread dE & 5259 keV \\ 
			average momentum P & 7.27 MeV/c \\ 
			transverse beam emittance eps x & 2.11E+04 pi mrad mm \\ 
%			correlated divergence cor x & 79.14 mrad \\ 
			transverse beam emittance eps y & 2.11E+04 pi mrad mm \\ 
%			correlated divergence cor y & 87.28 mrad \\ 
			longitudinal beam emittance eps z & 1.31E+05 pi mrad mm \\ 
%			correlated energy spread cor z & 1329 keV \\ 
%			emittance ration eps y/eps x & 0.93 \\ 
			\hline
		\end{tabular}
		\caption{Plasma lens design (more details, see \cite{niclas-bachelor}).}
		\label{tab:Tab1}
	\end{center}
\end{table}

\begin{figure}[H]
	\centering
    \includegraphics[scale=1]{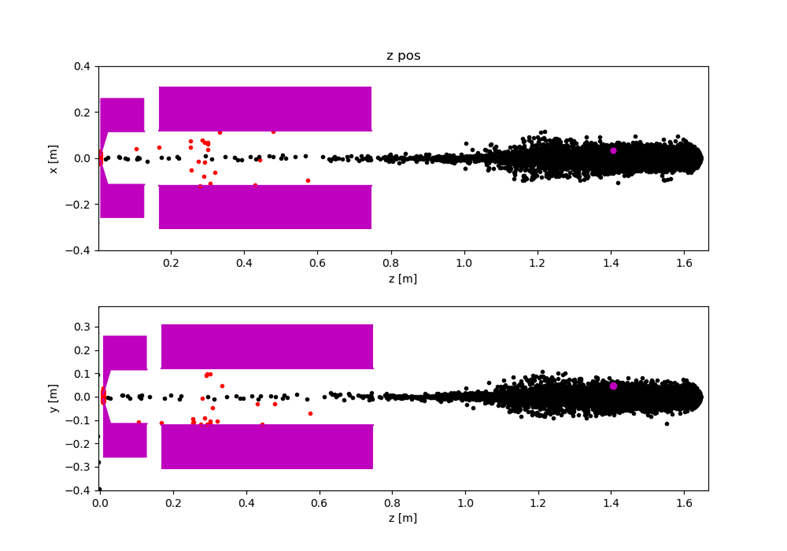}
    \caption{Zplot of a plasma lens.  Black dots belong to active particles and red dots belong to particles lost on the geometry.} \label{fig:zplot1}
\end{figure}

In table \ref{tab:Tab1} one can see the general beam statistics which consists of the spatial position, the beam size, the emittance in all three dimensions.\\
In Figures \ref{fig:zplot1}, we show the so-called z-plots of the discussed plasma lenses: the horizontal axis represents the z-position longitudinal to the beam and the vertical axis represents the transversal position of each particle. There black dots belong to active particles, red dots belong to particles lost on the geometry and blue dots denote particles that are set inactive via leaving the minimum z-position by travelling backwards.
\\ 

%\begin{figure}[H]
%	\centering
%    \includegraphics[scale=1]{bad_example_for_focused_bunch_2.png}
%    \caption{Zplot of another plasma lens. Black dots belong to active particles, red dots belong to particles lost on the geometry and blue dots denote %particles that are set inactive via leaving the minimum z-position by travelling backwards.}\label{fig:zplot2}
%\end{figure}

In the future we will further look into different plasma lens designs including also the damping ring acceptance and experimental feasibility. Also some parts of the RF cavities that are leading up to the damping ring are to be included, turning our simulations more sophisticated and still more appropriate,  cf.~\cite{niclas-bachelor,manuel-master}. However, already these first benchmarking simulations point to promising results for using plasma lenses as AMDs.

\section{Prospects and conclusions}

When compared to the quarter-wave transformer the plasma lens design in our simulation study 
has a similar and even better positron yield than the quarter-wave transformer,
since it is the effective field component which focuses the beam. Because of this the design of a plasma lens would be much smaller than a quarter-wave transformer. The total length of the quarter-wave transformer is  $12\,\mathrm{cm}$\cite{3}, whereas the plasma lens only 
needs $6\,\mathrm{cm}$. This study is still ongoing, however the results are very promising and should be pursued. The next steps are to include the damping ring acceptance and to address also
some technical aspects. One should note that plasma lenses are an active field right now and  many experiments are ongoing, as  FLASHForward
at DESY that will investigate, for instance, the stability of plasma for long pulse operation which is also of relevance for our study.
Therefore it can not yet been stated that the plasma lens is actually superior as OMD. Furthermore, one should keep in mind that
in addition to capture efficiency, 
also technical feasibility, reliability and costs are important factors as well. It is expected that not only the field of positron sources for high-energy colliders will benefit from  new developments in plasma lenses.

\section*{Acknowledgements}
GMP acknowledges by the Deutsche Forschungsgemeinschaft (DFG German Research Foundation) under Germany's Excellence Strategy --EXC 2121 "Quantum Universe" -- 390833306.

\end{document}